# FROM ELUSIVE TO UBIQUITOUS: UNDERSTANDING SMART CITIES

**Maria-Alexandra BARINA**
Politehnica University of Timișoara, România
vadasanmaria@gmail.com
**Gabriel BARINA**
Politehnica University of Timișoara, România
gabriel.barina@cs.upt.ro

**Abstract.** *Converting the city into a "smart" one is the emerging strategy of alleviating the problems generated by the rapid population growth in most urban areas, i.e., urbanisation. However, as the rate in which the different concepts of the smart city architecture are implemented is very high, academic research pertaining these advancements simply can not keep up. To improve the existing knowledge-base regarding the concept of smart cities, this paper reviews the existing definitions, as well as its architectures, based on an in-depth literature review of relevant studies and research fields alike. Additionally, we also provide our own definition of the smart city concept.*

**Nomenclature**
BD    Big Data
CC    Cloud Computing
CS    Cloud Storage
ICT   Information and Communications Technology
IoT   Internet of Things
SC    Smart City

## 1. Introduction

The concept of SC draws nowadays an ever-increasing attention from universities, research institutions and government. Whilst this concept first appeared in literature approximately 30 years ago, over time, it came to involve terms like IoT, BD, CC, CS, Energy, Transport and Mobility, Sensor Arrays, *etc*. As a result, due to this overlapping of scientific areas and terminologies in scholarly literature, a common understanding, as well as an exact definition for the concept of SC is still unclear [1].
It is with recent turnarounds, however, that scientists began focusing more on the concept of SC: overpopulation leads to environmental, economic, and social unsustainability of cities [2], increased traffic hinders optimal transportation [3] and energy consumption can not be maintained by means of renewable energy sources [4]. In other words, the rapid urbanisation of the world, albeit a symbol of evolution, brings about the difficulties associated with unsustainable energy consumption, saturated (public) transport networks, air and water contamination, toxic waste disposal, resource depletion, social disparity, public health decline, *etc*. [5], and at the same time, putting a considerable stress on urban systems functions and services [6]. As such, it became necessary that domains like IoT, BD, CS, *etc.* evolve, as it is with these domains that a city becomes smart and efficient [7].

Taking into account the existing bibliometric studies – while focusing solely on the overall (scientific) structure of this concept –, this paper aims at achieving the following objectives:
- Identifying currently accepted definitions regarding the concept of SC.
- Defining the building-blocks of a SC architecture.
- Proposing a new definition for the concept of SC.

To achieve the objectives above, we structured this paper as follows: in section 2, we discuss the terminology of the concept of SC throughout various academic literature and propose our own definition of SC, while in section 3 we describe the building-blocks of a SC architecture. In the last section, we draw the conclusions based on our findings.

## 2. Terminology

Throughout the last couple of decades, there have been different views regarding the umbrella-term of SC, or even its origin. According to selective literature, the concept of SC dates back to the '60s – at which point it was described as "cybernetically planned cities" – [12], although without any interest until the smart growth movement in the '90s [2], [13], [14]. During this period, the concept of SC came to encompass the idea of an increased urban efficiency with regard to natural resources, transportation and mobility, communications, *etc*. Indeed, the concept of SC focuses on efficiency, which in turn is based on the intelligent management of existing urban ICT devices.

At present, cities of all shapes and sizes are including propositions of SC in their urban sustainability policy. As a result, a proliferation in academic publications and other writings regarding the concept SCs [15] has been observed. This constant growth of literature articles over the course of the last two decades is highlighted in figure 1.

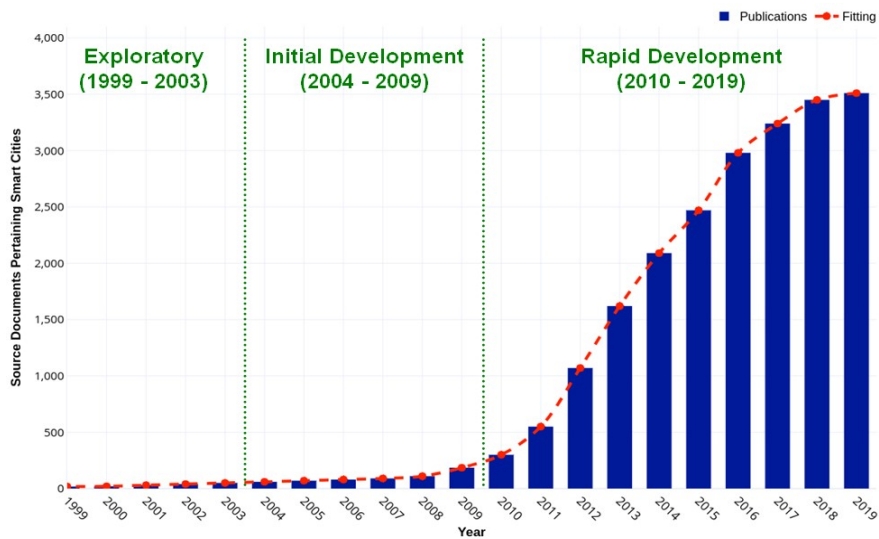

*Figure 1. Cumulative growth of source documents pertaining to Smart Cities over the course of two decades (1999 – 2019).*

Even so, there is no canonical or universally agreed upon definition for the concept of SC, and as such, it can be difficult at times to identify the true notation. In most cases, this concept is erroneously attributed to energy efficiency alone; even though efficiency is indeed an important aspect of SCs, it is merely a small portion of it. SC encompasses the entire human ecosystem [16], with focus on the current trend in the field of ICT to improve current applications, systems, architectures, technologies, *etc*.

Being a fairly new concept, SC requires further research in order to consolidate the technologies and ideas involved in the development of a SC. While in time this will aid in resolving all doubts with regards to the definition of the SC concept, we present the currently accepted (and summarised) ideas, namely:
- SCs are technologically advanced territories based on smart computing and a certain intellectual ability to improve our social, economical and technical evolution regarding smart infrastructures [17], [18], [19].
- SC is an umbrella-term for how advanced communication devices can be used to improve quality of life in any traditional city, while also promoting economic growth [20], [21], [22], [23], [24], [25].
- A SC uses sensor-networks, smart devices and ICT infrastructure in every aspect our daily life to promote resource productions and optimise its consumption [26], [27], [28], [29], [30], [31].
- A SC represents an efficient, technologically advanced, and sustainable city, an urban innovation ecosystem [32], [33], [34].
- SC refers to a local entity – *e.g., city, district, region, etc.* – which employs a comprehensive computerised approach to gather and analyse a large quantity of real-time data and create added value in the process [19], [29], [30].

Taking into consideration the above definitions regarding SCs, we can see how they all focus on the integration, as well as the constant development of ICT, BD, IoT theories, methodologies and devices, in order to offer citizens an improved lifestyle. To this end, we, the authors, have defined the concept of SC as being an urban development with strong focus on the relation of its citizens and ICT, both variable and fixed.

## 3. Smart city architecture conceptualisation

To obtain a distributed and autonomous SC, all new devices come with embedded electronics in order to control, as well as enable them to link with other devices. To this end, the ICT infrastructure is used to obtain a grid-wide coordinated monitoring system by harnessing modern communication and information technologies. However, such an infrastructure must also be capable of offering spontaneous, continuous and reliable bi-directional communication between each interconnected device within the respective grid [35]. Consequently, such an infrastructure assumes transmitting, processing and storing vast amount of data at an increased frequency, thus heavily relying on Cloud Storage, Cloud Computing and Big Data [36], [37]. Indeed, by using such services, one can expect important advantages, like cost savings, improved city management, reliable prediction, [16] *etc*.

### 3.1 From a city to a smart city

To make a city smart, we need to take into consideration certain functionalities, basic services and technologies, including sensor-arrays and devices, for information gathering and analysis. Hence, existing city platforms are examined to the point of identifying and improving existing defects, as well as to achieve the technological level needed for a SC; after all, this technological level also creates public value, by focusing all existing projects and initiatives towards its citizens [38].

SC technology can also be beneficial for the daily life as well by identifying the domains (and their individual services) that are often offered to society in a structured way [2]. As such, the existing literature defines the following domains of a SC: natural resources, mobility, construction, daily life, government, economy and society; these, along with the individual services offered, are presented in figure 2. However, due to the rapid evolution of technology, these services are being constantly extended, in order to support the existing set of technological solutions.

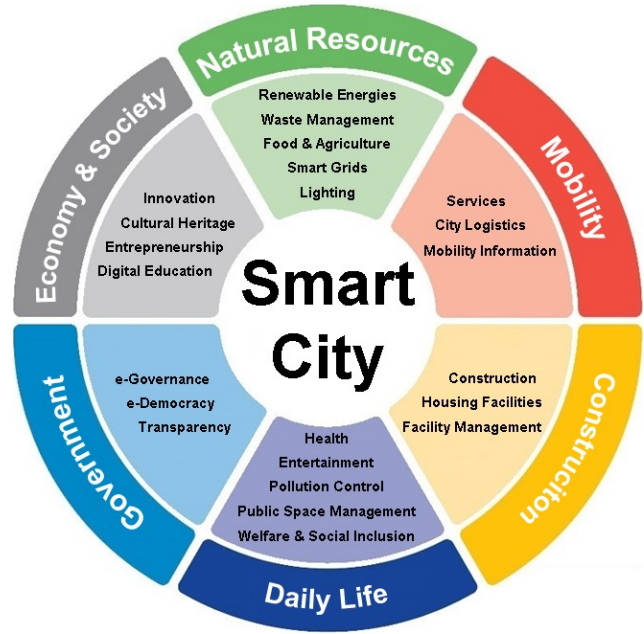

*Figure 2. The classification of the Smart City domains and the corresponding delivered services.*

### 3.2 Information and Communications Technology

Smart cities are made up of multi-layered ICT structures, able to handle multi-directional flows of information between crucial sectors, such as transportation and mobility, resources, economy and society, *etc*.

From a technical standpoint, ICT includes both hardware and software components [39]; the hardware elements includes sensor arrays (*e.g., GPS, IR, RFID, wearable devices etc.*), computers, smartphones, telecommunication infrastructure, cloud computing, *etc.*, whereas software elements – besides the applications needed for running the mentioned hardware systems –, includes applications for analysing Big Data (*e.g., data-mining, machine-learning, (real-time) simulation and analysis, etc.)*, software needed for database integration, simulation application software, communication protocols, *etc*.

Overall, there are two popular approaches to a SC with the integration of ICT infrastructure [40], [41]:

- The ICT-oriented (technological) approach: spanning over several urban domains, this approach focuses on integrating ICT in the existing hard infrastructure, like transportation, facilities, energy–, and communication–networks, or other (physical) objects.
- The population-oriented approach: focusing on the integration of ICT into wearable devices, soft infrastructure and people, it refers to the context of its daily usage: smart transportation, smart traffic, smart safety, smart energy, smart healthcare, smart equity, smart education, *etc*.

### 3.3 Big data

As it stands, 90% of the world's digitised data was obtained over the last couple of years, sourced from the plethora of IoT sources of communicated data [36] within SCs. As a result,

BD is all about dealing with large volumes of information, which is beyond the capabilities of regular databases to be retrieved, stored, managed and analysed [42], [43].

The concept of SC is tightly coupled with the BD, IoT, as well as on-site sensor arrays, leading to an informed and data-driven approach [44], [45]. Nonetheless, despite their strong and well-documented association, it is not a contemporary concept [46]; instead – from a historical point of view –, the idea of large-scale data analysis of cities is a century-old concept, popularly embraced by both planners and engineers alike [47], [48]. Nowadays, the definition of BD has been broadened; consequently, what began with only a large data-set, now encompasses the data, its analysis and the subsequent interpretation and presentation [16], [49].

Currently, BD applications are applied in several sectors of SCs [50] through the ICT infrastructure [51]. Local governments have started applying BD in order to fully support both the development and sustainability of such cities; this allows them to maintain standards, principles as well as requirements needed in order to offer sustainability, mobility, enhanced quality of life and resource management [51], [52].

### 3.4 Cloud Computing

Due to the inherent data-driven nature of the SC architecture, all connected devices transmit large quantities of data to each other. However, since the number of devices may vary – or the volume of generated data –, the underlying infrastructure must also be capable of adapting to current requirements. As such, in order to provide a flexible way of adapting storage space, as well as computing power, the concept of SC is closely related to that of the CC [53]. Specifically, CC refers to applications, services, as well as hardware/software systems capable of adapting computation resources without having the actual physical infrastructure to provide them, transforming capital expenditures into operational costs [16], [54].

The application of CC in SCs has multiple advantages. For instance, running BD applications on a CC infrastructure allows for real-time monitoring, processing, analysis, interpretation and reaction to processes, activities, events, behaviours, environmental states, socio-economic patterns, *etc.*, and taking action accordingly. Oftentimes, this means that CC is used for greatly improving energy usage, street and traffic lighting, transportation and mobility, healthcare, education, safety, [6], [40] *etc*. Therefore, in [55] Lu *et al.* propose a unique, CC-based framework for multi-scale climate data analytics, capable of differentiating between urban climate targets and opportunities offered by the ICT solutions [6], [56].

As cloud services constantly improve to store, analyse and extract relevant information from the raw data obtained by IoT devices connected to it, so do these devices evolve as well. Capable of performing relevant storage and computational tasks directly on-site [57], by combining CC with advanced IoT devices can indeed offer for any SC dynamic computing capabilities, scalable based on varying demands of responsiveness and complexity [58], [59]. To this end, Carnevale *et al.* present a state-of-the-art design for cloud architecture, capable of enabling smart IoT applications [60], validating it against a real-world smart campus test-bed. Consequently, they showcase a novel cloud-assisted, cyber-physical-social framework capable of migrating tasks from cloud services to localised devices, considerably improving latency, as well as real-time view of events.

### 4. Closing thoughts

All concepts presented throughout this paper are indeed important and are encompassed by the overarching concept of SC. As a result, their integration into SC applications is mandatory, in order to achieve sustainability, enhanced mobility and transportation, effective governance, improved quality of life, intelligent city resources management, *etc*.

This review highlights the main concepts of the SC research domain, in which technology plays a crucial role in shaping the contemporary urban scenery, by empowering users and citizens alike in sustainable urban planning, intelligent behaviour and decision-making. The relevant theories, concepts and technologies are highlighted and elaborated on in regards with the topic of smart sustainable cities. However, as presented in this paper, most approaches towards a SC are inherently difficult in implementation; building and successfully employing ICTs and their integration in urban areas requires addressing challenges, following design models, employing computer simulation, having well trained human resources and being supported by the government entities. With all these in place, making a city smart is indeed achievable, leaving room for further enhancements as well.

By using a bibliometric approach of the SC literature over the last two decades, we have highlighted the emergence of the concept of SCs, the current state of this research field and the growing number of contributions. To this end, this paper contributes to the leveling of the state-of-the-art in SC architecture by highlighting its main components.